# Molecular frequency reference at 1.56µm using a $^{12}C^{16}O$ overtone transition with the noise-immune cavity-enhanced optical heterodyne molecular spectroscopy (NICE-OHMS) method

SHAILENDHAR SARAF*, PAUL BERCEAU, ALBERTO STOCHINO, ROBERT BYER, AND JOHN LIPA

*Hansen Experimental Physics Laboratory, Stanford University, Stanford, CA 94305 USA*
*Corresponding author: saraf@stanford.edu*

**We report on a molecular clock based on the interrogation of the 3ν rotational-vibrational combination band at 1563 nm of carbon monoxide $^{12}C^{16}O$. The laser stabilization scheme is based on the NICE-OHMS technique in frequency modulation (FM) saturation spectroscopy with a high-finesse ultra-low expansion (ULE) glass optical cavity using CO as the molecular reference for long-term stabilization of the cavity resonance. We report an Allan deviation of $1.8 \times 10^{-12}$ at 1 second that improves to $\sim 3.5 \times 10^{-14}$ with 1000 seconds of averaging. © 2015 Optical Society of America**

http://dx.doi.org/10.1364/OL.99.099999

There are several applications for compact stable frequency standards operating in space such as advanced GNSS systems, tests of fundamental physics, formation flying, and gravitational wave detectors [1]. The STAR (Space and Time A-symmetry Research) experiment [2] is another example of the need for a space-based ultra-stable clock for rod-clock type experiments designed to test Lorentz invariance. Molecular clocks based on modulation transfer spectroscopy (MTS) using an electronic transition of $^{127}I_2$ interrogated at 532nm have been developed for space applications [3]. However, these clocks need precise pointing control, have magnetic sensitivity and are limited to $10^{-14}$ frequency stability at short timescales [4]. Our objective is to advance the development of a laser stabilization scheme that has very low noise into a unit that is low power and compact, and can be used in space for missions requiring extreme performance in optical frequency stabilization. Insensitivity to DC and AC magnetic fields is required for these clocks, especially payloads operating in a Low Earth Orbit. Our device operates near 1563 nm using low-pressure CO gas as a molecular reference and a spectroscopic technique that has been demonstrated using gases such as $C_2HD$, $C_2H_2$, $H_2O$ and $CO_2$ [5-9]. Its ultimate performance as a stabilization system using CO, based on calculation [10], is expected to approach an Allan deviation of $\sim 2 \times 10^{-16}$ at 1 sec, not far from that of ultra-cold neutral atom clocks that require multiple lasers and an extensive amount of equipment. The lighter and simpler package that we are developing has the potential of yielding a more practical ultra-stable reference for space where mass, power and complexity are critical issues. Another reason for choosing CO was to operate near the 1550 nm telecom band for which many components are readily available.

Molecular overtones provide many rotation-vibrational comb-like resonances that cover a broad spectrum in the visible and near-IR regions, while the linewidths potentially are in the sub-kilohertz domain that is characteristic of the decay rates of fundamental molecular stretch resonances. For frequency stabilization purposes, we need to observe saturated absorption resonances to achieve the necessary narrow sub-Doppler linewidths that are inevitably broadened well beyond the decay rate limit. The weak absorption of these overtone transitions requires high optical powers to achieve saturation. For example, a typical cavity power needed for $C_2HD$ $(v_2+3v_3)P(5)$ transition is >100 W at $\sim 1$ mTorr for the 1064nm transitions [5]. For the 1560 nm CO transitions, the cavity power needed for saturation is $\sim 2W$ at 10 mTorr. Thus for these transitions we are led to consider cavities of high finesse to yield large power buildup inside the cavity, giving good saturation parameters without requiring a large transmitted power to be accepted by the detector. However, the cavity with high finesse naturally becomes an efficient laser-frequency discriminator, leading to conversion of residual laser-frequency noise into amplitude noise [6].

In conjunction with saturation spectroscopy, FM

spectroscopy avoids excessive amplitude noise, by performing modulation and detection of the signal in a frequency region where noise processes of whatever origin no longer possess strong power density. The NICE-OHMS technique [5-9,11-13] captures the advantages of both FM detection and intra-cavity absorption enhancement while eliminating sensitivity to the laser/cavity relative frequency noise. The frequency offset of the modulation sidebands for FM detection is actively controlled to fall exactly on axial cavity resonances, shifted by one or more cavity free spectral range (FSR). Therefore, for small frequency shifts of the laser and its sidebands about the exact cavity modes, all experience the same small amplitude jitter and phase shifts for their internal fields. Thus the transmitted beam remains purely FM despite small power fluctuations on the detector, and no unwanted detected noise results from small laser–cavity detunings at the modulation frequency, thus rendering the spectroscopic method as being frequency-noise immune [6].

Here we report the first NICE-OHMS based molecular clock at the telecom wavelength band interrogating a narrow overtone transition of carbon monoxide. CO is an attractive molecule for metrology as it has several rotational-vibrational transitions and associated overtone bands that can be interrogated with commercially available near-IR lasers. A total rotational-vibrational energy calculation yields a spectrum for the diatomic CO molecule with the R and P branches on each side of the band origin [14]. The strongest transition for the 2nd overtone ($\Delta v =3$) is the R(7) line at 1568.038nm. In the early stage of our experiment, the laser source customized to interrogate this transition was noisy causing us to change to the weaker R(14) line at 1563.618nm. Key spectroscopic parameters for this line [8,10] are $\alpha_m$, the small-signal absorption coefficient of $2.2\times10^{-8}$ cm$^{-1}$/mTorr, the pressure-broadening coefficient of 5.11 kHz/ mTorr and the pressure shift of -213 Hz/ mTorr [14]. The Doppler broadened linewidth at 25 °C is 443 MHz (FWHM). Band origin for the 3v band is at 6429.9 cm$^{-1}$. Dipole moment for this transition [15,16] is 0.41 mDebye. The saturation power based on this calculated dipole moment at the operating gas pressure of 9.5 mTorr is 1.95 W. Frequency shifts from Zeeman splitting are < 4 MHz/Tesla [17].

At the heart of the experiment is a 25.6 cm optical cavity-- formed by a ULE spacer between a pair of flat and 50 cm radius-of-curvature (ROC) high-finesse super-mirrors. The cavity is installed on a V-block inside an aluminum can that can be filled with CO gas at a desired pressure, usually set at 9.5 mTorr. The fused-silica substrate matched super-mirrors (Advanced Thin Films) are attached to PZTs (Piezomechanik, 150/14-10/12) that are mounted on the ULE spacer. The cavity FSR is 585 MHz, the spot sizes (radii) on the flat and curved mirror are calculated at 353 μm and 504 μm, respectively. The average spot radius for the Gaussian TEM$_{00}$ beam in the cavity for estimating the transit-time broadening is calculated at 411 μm. The empty-cavity finesse using a ring-down measurement is 170,000, yielding a round-trip cavity loss of 37 ppm. Mirror loss is estimated at 8 ppm, implying the transmittance of the output mirror to be 10.5 ppm. Transit time broadening for a spot of diameter of 822 μm and the most probable speed of CO molecules at room temperature is calculated [6] at 257 kHz. At 9.5 mTorr the molecular linewidth becomes 308 kHz. Since we operate with a saturation parameter $s = 0.8$, the power broadened linewidth is 413 kHz. The single-pass small-signal integrated absorption $\alpha_0 L = \alpha_m L p$, where $p$ is the gas pressure, is calculated to be 5.5 ppm for the R(14) line at 9.5 mTorr. The absolute level $\Delta\alpha L$ of saturated absorption by the CO molecules inside the cavity for low levels of the saturation parameter $s$ is given by [6]

$$\Delta\alpha L = \left(\frac{\alpha_0 L}{\sqrt{1+s}} - \frac{\alpha_0 L}{\sqrt{1+2s}}\right). \quad (1)$$

where $L$ is the length of the cavity. Therefore the absolute level of saturated absorption in our CO cavity based on a measured peak contrast of 12.5% is 0.69 ppm. We note that Axner's formula [11] gives a 14% higher value. This number divided by the measured S/N from the NICE-OHMS detection system [7] will yield the noise-equivalent species detection sensitivity of the spectrometer and is estimated below.

The non-resonant single-pass cavity gas absorption plus mirror loss for $s = 0.8$ is 12.1 ppm off the sub-Doppler peak. However, at line center with 2-beam saturation, the single-pass absorption is 11.4 ppm. The resonant cavity Finesse based on these absorption numbers is calculated at 143,000 yielding a linewidth of 4.1 kHz. The cavity power buildup is 21,800. The input power for optimal saturation is about 90 μW, assuming optimal coupling. Transmission coefficients off-resonance and on-resonance are 21.7% and 23%, respectively, and the corresponding transmitted powers are 19.4 μW and 20.6 μW. Therefore, the signal power is 1.2 μW and assuming a photodetector efficiency of 0.95 mA/mW, we obtain a S/N of $4.7\times10^5$ at 1 second, estimating the noise as in [7]. The corresponding fractional frequency stability $\delta\nu/\nu$, is given by

$$\left.\frac{\delta\nu}{\nu}\right|_{1s} = \frac{\Delta\nu}{\nu}\frac{1}{S/N}, \quad (2)$$

where $\nu$ is the laser frequency and $\Delta\nu$ is the molecular transition linewidth. The molecular linewidth divided by the S/N is 0.87 Hz yielding a fractional frequency stability of $4.5\times10^{-15}$ at 1 second. This will be a practical limit of the performance of our setup. On the other hand, from a detection sensitivity perspective, the shot-noise-limited sensitivity of our spectrometer is given by [6]

$$(\alpha L)_{\min} = \frac{\pi}{F}\sqrt{\frac{h\gamma B}{\eta P_0}}\frac{1}{J_0(\beta)J_1(\beta)}, \quad (3)$$

where $F$ is the cavity finesse, $h$ is Planck's constant, $B$ is the detection bandwidth, $\eta$ is the detector's quantum efficiency, $\gamma$ is the carrier frequency, $P_0$ is the light power on the transmission detector and $\beta$ is the modulation index. Our CO spectrometer operated at the optimal $\beta = 1.1$ and $\eta = 0.95$ mA/mW, detection BW of 1/2 Hz and $P_0 = 20.6$ μW should yield a shot-noise-limited species detection sensitivity $(\alpha L)_{\min}$ of $1.75\times10^{-12}$ at 1 second.

Figure 1 shows our experimental setup. The light source is a 12 mW single-frequency, narrow-linewidth external-cavity diode laser [18] at a custom wavelength of 1563.62 nm from Redfern Integrated Optics (RIO). The laser is first locked to a low-finesse (~ 6000) ULE pre-stabilization cavity temperature controlled in a vacuum using a free-space double pass acousto-optic modulator (AOM), as shown at the top of the figure. The free running fractional stability is ~ $2\times10^{-9}$ at 1 sec which falls to ~ $2\times10^{-10}$ after locking. The AOM suppresses the high frequency noise of the laser while a piezoelectric (PZT) transducer on a cavity mirror is adjusted to allow the cavity to track the average laser frequency. The AOM is driven by a wide bandwidth, low-noise voltage controlled oscillator with a null frequency of 200 MHz.

The laser beam is then amplified to 150 mW by a fiber amplifier (FA, PriTel, PMFA-30) to compensate for the relatively high insertion loss in the AOM and EOM. We then lock the laser to our ultra-high finesse Fabry-Perot cavity with the Pound-Drever-Hall (PDH) method [19]. To lock the laser on the cavity, a first pair of sidebands ($f_{PDH}$ = 30 MHz) with a modulation index $\beta_1 \sim 0.4$ are added to the beam using an electro-optic modulator (EOM), and heterodyne detection of the beam reflected by the cavity is performed by the photodiode PD1 (Newfocus, 1611). This produces an error signal that is processed to drive a fiber-coupled double-pass AOM (Brimrose, IPF-300-100-1550-2FP). Our servo bandwidth is 150 kHz and its gain is sufficient to provide a low frequency-noise density of 0.05 Hz/$\sqrt{Hz}$ corresponding to a 51 mHz laser linewidth relative to the cavity [20]. The input power to the cavity is set at 1.1 mW and is stabilized using a small pickoff photodiode PD2 and a servo loop whose error signal drives the RF power sent to the AOM. The fractional power noise is typically < $10^{-3}$. We also add a second pair of sidebands at $f_{FSR}$ = 585 MHz with a modulation index $\beta_2 \sim 0.45$ using a signal generator (Stanford Research Systems, SG382) with an external modulation input and lock them to the FSR of the cavity by implementing the DeVoe-Brewer technique [21]. The heterodyne detection is performed by the photodiode PD3 at the frequency $f_{FSR}$ - $f_{PDH}$ = 555 MHz, and the feedback signal is sent to the external modulation port of the signal generator. The detection phase of the error signal is optimized using a phase delay $\Delta\Phi_2$ between the signal generator and the bandpass filtered signal from the reflection photodetector. The bandwidth of the FSR servo loop is ~10 kHz. The carrier power coupled into the cavity is estimated based on the modulation indices of the PDH and FSR sidebands and the reflected power. The coupling efficiency is about 12% and the circulating cavity power corresponds to a saturation parameter s ~ 0.8. Once the carrier and FSR-sidebands are resonant with the cavity we perform heterodyne frequency detection in transmission with the photodiode PD3 (Newfocus, 1611) at $f_{FSR}$. The carrier is brought onto the resonance of the CO line by setting the DC voltage on the input cavity mirror PZT. The carrier thus strongly interacts with the gas, unlike its sidebands, too far from resonance, which become local oscillators used for the heterodyne detection. We band-pass filter the $f_{FSR}$ component of the signal from the transmission photodetector and mix it with the 585 MHz source. The resulting error signal is optimized by setting the detection phase to dispersion using the phase delay $\Delta\Phi_1$ between the two RF signals.

Fig. 1. (color online) Simplified layout of the experimental setup showing the light source and the NICE-OHMS stabilization system: AOM, acousto-optic modulator; EOM, electro-optic modulator; FA, fiber amplifier; PBS, polarizing beam splitter; FR, Faraday rotator; PID, servo control stages; $f_{PDH}$, PDH modulation frequency; $f_{FSR}$, free spectral range lock frequency; $f_{dith}$, dither frequency; PD1, reflection photodiode; PD2, input power photodiode; PD3, transmission photodiode; $\Delta\Phi_{1,2}$, phase shifters; PZT, piezoelectric transducer; LPF, low-pass filter; BPF, band-pass filter; **X**, double balanced mixer.

To improve the S/N and baseline stability, we dither the cavity length (on the PZT of the output mirror) at the frequency $f_{dith}$ = 149 Hz that corresponds to a local minimum in the observed power spectral density of the RF balanced mixer. The frequency deviation of the applied dither is ~ 100 kHz. The demodulated low-pass filtered error signal is the input to a lock-in amplifier (SRS model SR830) for derivative locking [22] and the 2f signal is fed back to the cavity input mirror PZT.

Fig. 2. Top: Sub-Doppler transmission peak (in grey) and demodulated NICE-OHMS signal (in black). Center and Bottom: Dither lock-in detection of the NICE-OHMS signal at the first and second harmonics. A slight asymmetry is caused by the non-linear response of the piezos.

The upper plot in Figure 2 shows the transmitted sub-Doppler peak in gray along with the demodulated NICE-OHMS signal as a function of the frequency detuning from line center (in black). The linewidth of the saturated absorption peak was measured to be ~ 400 kHz, mainly due to the transit time broadening and pressure broadening as calculated above. The demodulated signal shows a S/N of 458 at 1 sec. The center and lower plots show the demodulated signals at the 1f and 2f harmonics of the dithered NICE-OHMS signal. The time constant of the lock-in amplifier was set at 30 ms and the measured S/N for the 1$f$ and 2$f$ harmonics was 216 and 222, respectively. Therefore, the S/N at 1 sec is calculated at 1249 and 1280 respectively. The fractional stability of our frequency standard is the transit-time, pressure and power broadened molecular linewidth divided by the S/N of the 2$f$ signal at 1 sec. Therefore, based on the measured S/N, the fractional frequency stability is estimated to be $1.63\times10^{-12}$ at 1 sec. Additionally, the species detection sensitivity based on the absolute level of saturated absorption calculated earlier and the S/N of the 1$f$ detected signal works out to $5.5\times10^{-10}$ at 1 sec.

We evaluated the frequency stability of our NICE-OHMS based molecular clock system by comparing it to a frequency comb. We pick off part of the incident beam after the fiber-coupled AOM, and generate a beat signal with the frequency comb (Menlo Systems, FC1550). The 250 MHz repetition rate of the comb is stabilized by locking one tooth to a temperature-stabilized Fabry-Perot cavity ($F$ = 130,000) in vacuum with a measured drift < 1 Hz/sec. The cavity fractional frequency instability is below $10^{-14}$ at 1 sec. The frequency of the RF beatnote is logged with a 10-digit counter (Keysight, 53230A). The offset frequency of the comb is 10 MHz and it is stabilized with a hydrogen maser (T4Science, iMaser3000) that has a measured fractional frequency instability of $10^{-13}/\sqrt{\tau}$, where $\tau$ is the averaging time. Figure 3 shows the observed fractional frequency stability of the CO frequency standard with an Allan deviation of $1.8 \times 10^{-12}$ at 1 sec, averaging down approximately as $1/\sqrt{\tau}$ to $\sim3.5\times10^{-14}$ at 1000 sec and still trending lower. The upper time limit of the plot is 2048 sec, set primarily by the operation of the comb.

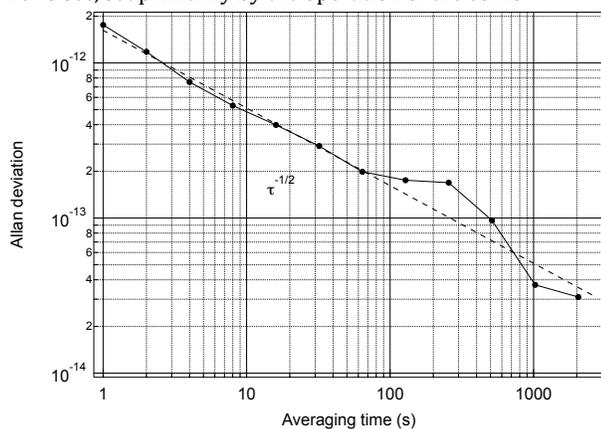

Fig. 3. Allan deviation versus averaging time of the beat-note between the NICE-OHMS frequency stabilized system and a cavity-stabilized frequency comb.

The bump in the plot appears to be caused by the air conditioning cycle in the laboratory. The value at 1 sec is consistent with the number calculated above by considering the measured S/N from Fig. 2. The Fabry-Perot cavity shows lower frequency instability at short times, mainly limited by the noise of the PZTs holding the cavity mirrors.

It is clear that our detection system has good sensitivity, but has not reached the shot-noise floor. Potential S/N limitations arise from technical noise, frequency jitter, scattered light, residual amplitude modulation (RAM) and radiated coupling from the relatively high 585 MHz FSR frequency. We are working to improve the S/N of our system by using isolators and resonant photodetectors in transmission, optimization and shielding of the electronics layout. In order to reduce piezo noise, we plan to use shorter piezos and also attempt to place piezo rings around the ULE cavity for length control. We are customizing the laser source to interrogate the stronger R(7) absorption line at 1568.038 nm that should roughly double the molecular contrast. We are reducing the transit-time line broadening through an increase in the cavity mode size by the use of high ROC mirrors (> 5 m). We are in the process of building a second CO spectrometer for direct frequency stability measurements and optimization of the spectroscopic, optical and electronic parameters in a systematic manner to approach the shot-noise limit of the frequency standard.

In summary, we have demonstrated the first laser frequency stabilization scheme with a molecular $^{12}C^{16}O$ overtone transition, using the NICE-OHMS method. The measured frequency instability is $1.8\times10^{-12}$ at 1 second and $\sim3.5\times10^{-14}$ at 1000 seconds.

**Funding.** NASA Strategic Astrophysics Technology (SAT) program "Physics of the Cosmos" (PCOS) under grant #NNX13AC91G.

**Acknowledgements**. We thank J.L. Hall for useful discussions and use of his modeling spreadsheet. We also thank L. Hollberg at Stanford University for providing access to the frequency comb for clock stability measurements, and S. Tan, K. Urbanek and S. Wang for help with the optics and electronics.

## FULL REFERENCES

1. K. Danzman and the LISA study team, "LISA: laser interferometer space antenna for gravitational wave measurements," Classical and Quantum Gravity, **13**, A247 (1996).
2. J. Lipa, S. Buchman, S. Saraf, J. Zhou, A. Alfauwaz, J. Conklin, G.D. Cutler, and R.L. Byer, "Prospects for an advanced Kennedy-Thorndike experiment in low Earth orbit," *arXiv*:1203.3914v1 [gr-qc] (2012).
3. T. Schuldt, S. Saraf, A. Stochino, K. Doringshoff, S. Buchman, G. D. Cutler, J. Lipa, S. Tan, J. Hanson, B. Jaroux, C. Braxmaier, N. Gurlebeck, S. Herrmann, C.Lammerzahl, A. Peters, A. Alfauwaz, A. Alhussien, B. Alsuwaidan, T. Al Saud, H. Dittus, U. Johann, S.P. Worden, and R.L. Byer, "mSTAR: Testing Special Relativity in Space Using High Performance Optical Frequency References," IEEE International Frequency Control Symposium & European Frequency and Time Forum Proceedings, Denver, Colorado, April 2015
4. E. J. Zang, J. P. Cao, Y. Li, C. Y. Li, Y. K. Deng, and C. Q. Gao, "Realization of four-pass I2 absorption cell in 532-nm optical frequency standard," IEEE Trans. Instrum. Meas. **56**, 673-676 (2007).
5. J. Ye, L. S. Ma, and J. L. Hall, ''Sub-Doppler optical frequency reference at 1.064 mm by means of ultrasensitive cavity-enhanced frequency modulation spectroscopy of a C2HD overtone transition,'' Opt. Lett. **21**, 1000–1002 (1996).
6. L. S. Ma, J. Ye, P. Dubé, and J. L. Hall, ''Ultrasensitive frequency-modulation spectroscopy enhanced by a high finesse optical cavity: theory and application to overtone transitions of C2H2 and C2HD,'' J. Opt. Soc. Am. B **16**, 2255–2268 (1999).
7. J. Ye, L. S. Ma, and J. L. Hall, ''Ultrasensitive detections in atomic and molecular physics: demonstration in molecular overtone spectroscopy,'' J. Opt. Soc. Am. B **15**, 6–15 (1998).
8. A. A. Madej, A. J. Alcock, A. Czajkowski, J. E. Bernard, and S. Chepurov, "Accurate absolute reference frequencies from 1511 to 1545 nm of the v1+v3 band of 12C2H2 determined with laser frequency comb interval measurements," J. Opt. Soc. Am. B **23**(10), 2200–2208 (2006).
9. H. Dinesan, E. Fasci, A. Castrillo, and L. Gianfrani, "Absolute frequency stabilization of an extended-cavity diode laser by means of noise-immune cavity-enhanced optical heterodyne molecular spectroscopy," Opt. Lett. **39**, 2198-2201 (2014).
10. J. L. Hall, private communication.
11. O. Axner, W. Ma, and A. Foltynowicz, "Sub-Doppler dispersion and noise-immune cavity enhanced optical heterodyne molecular spectroscopy revised," J. Opt. Soc. Am. B **25**, 1166–1177 (2008).
12. A. Foltynowicz, F. M. Schmidt, W. Ma, and O. Axner, "Noise-immune cavity-enhanced optical heterodyne molecular spectroscopy: current status and future potential," Appl. Phys. B **92**, 313–326 (2008).
13. N. J. Van Leeuwen and A. C. Wilson, "Measurement of pressure-broadened, ultraweak transitions with noise-immune cavity-enhanced optical heterodyne molecular spectroscopy," J. Opt. Soc. Am. B **21**, 1713-1721 (2004).
14. W. C. Swann, S. L. Gilbert, "Pressure-induced shift and broadening of 1560–1630-nm carbon monoxide wavelength-calibration lines," J. Opt. Soc. Am. B **19,** 2461-2467 (2002).
15. J. Henningsen, H. Simonsen, T. Møgelberg, and E. Trudso," The 0→ 3 Overtone Band of CO: Precise Linestrengths and Broadening Parameters," Journal of Molecular Spectroscopy **193,** 354–362 (1999).
16. L. S. Rothman, I. E. Gordon, Y. Babikov, A. Barbe, D. Chris Benner, P. F. Bernath, "The HITRAN2012 molecular spectroscopic database," *JQSRT* **130**, 4-50 (2013).
17. P. H. Krupenie, "The Band Spectrum of Carbon Monoxide," NSRDS-NBS 5, National Standard Reference Data Series, NBS- 5 (1966).
18. Kenji Numata, Jordan Camp, Michael A. Krainak, and Lew Stolpner, "Performance of planar-waveguide external cavity laser for precision measurements," Opt. Express **18**, 22781-22788 (2010).
19. R. W. P. Drever, J. L. Hall, F. V. Kowalski, J. Hough, G. M. Ford, A. J. Munley, and H. Ward, "Laser phase and frequency stabilization using an optical resonator," Appl. Phys. B **31**, 97–105 (1983).
20. Gianni Di Domenico, Stéphane Schilt, and Pierre Thomann, "Simple approach to the relation between laser frequency noise and laser line shape," Appl. Opt. **49**, 4801-4807 (2010).
21. R. G. DeVoe and R. G. Brewer, "Laser frequency division and stabilization," Phys. Rev. A **30**, 2827–2829 (1984).
22. H. Wahlquist," Modulation Broadening of Unsaturated Lorentzian Lines," J. Chem. Phys. **35**, 5, 1708-1710 (1961).